# The Chebyshev Polynomials Of The First Kind For Analysis Rates Shares Of Enterprises

Sergey Yekimov*

Czech University of Life Sciences Prague, 165 00 Praha – Suchdol, Czech Republic; yekimov@pef.czu.cz
* Corresponding author: yekimov@pef.czu.cz

**Abstract:** Chebyshev polynomials of the first kind have long been used to approximate experimental data in solving various technical problems. Within the framework of this study, the dynamics of shares of eight Czech enterprises was analyzed by the Chebyshev polynomial decomposition: CEZ A.S. (CEZP), Colt CZ Group SE (CZG), Erste Bank (ERST), Komercni Banka (BKOM), Moneta Money Bank A.S. (MONET), Photon (PENP), Vienna insurance group (VIGR) in 2021.

An investor, when making a decision to purchase a security , is guided largely by an heuristic approach . And variance and correlation are not observed by human senses.

The vectors of decomposition of time series of exchange values of securities allow analyzing the dynamics of exchange values of securities more effectively if their dynamics does not correspond to the normal distribution law.

The proposed model allows analyzing the dynamics of the exchange value of a securities portfolio without calculating variance and correlation.

This model can be useful if the dynamics of the exchange values of securities does not obey, due to certain circumstances, the normal law of distribution.

**Keywords:** Chebyshev polynomials of the first kind, securities portfolio, heuristic approach, exchange rate value of a security, variance, correlation.

## 1. Introduction

The process of development of the world economy is closely correlated with the development of securities markets. The securities market has long been the locomotive of the global economy. The integration of national securities markets into a single, global market allows investors to carry out their investment activities regardless of their location.

The formation of a single global securities market became possible due to the development of international banking settlements, as well as the standardization of regulation on national stock markets.

The authors (Morhachov, Illia, 2022), the main purpose of financial investments is the placement of temporarily free financial resources to generate investment income. Modern investors in most cases choose several different financial instruments to make financial investments.

According to (Šálková, Daniela et al. ,2017), the most common objects of financial investments are (Figure 1):

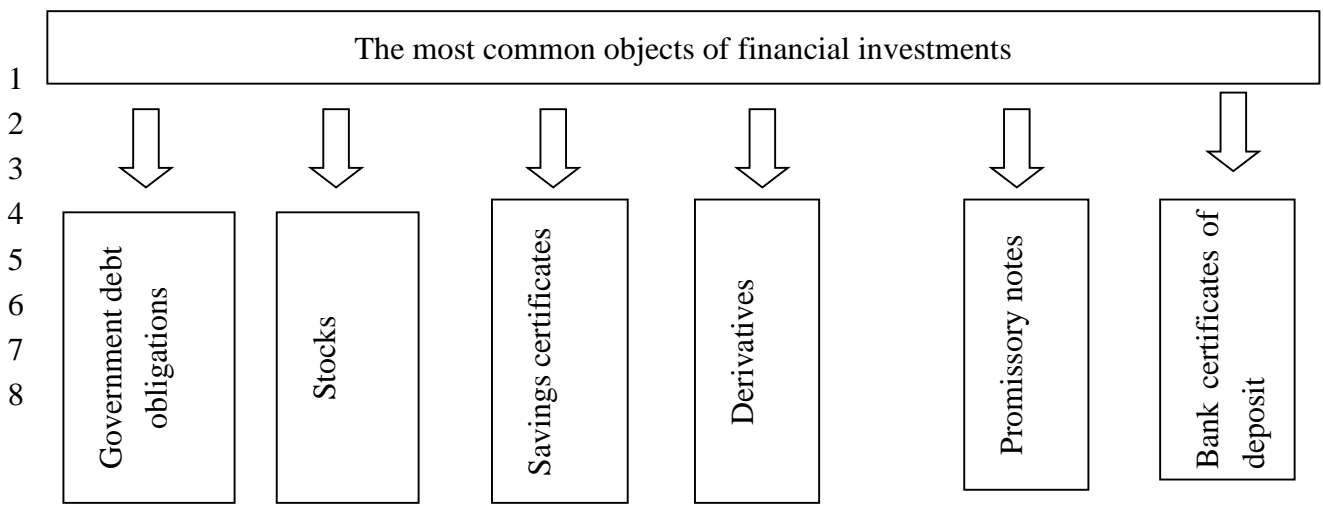

**Figure 1.** The most common objects of financial investments

In views to (Myroshnychenko, Ihnat et al. ,2022), in order to form an effective investment portfolio, investors diversify their activities not only between different types of securities, but also individual regions or countries.

The authors (Dvořák, Marek et al.,2021) note that the investment portfolio can be characterized by:

1) The level of liquidity.

2) The level of investment risks.

3) The level of liquidity.

4) The size of the investment period.

5) Stability of the investment portfolio structure.

According to (Morhachov, Illia, 2022) the following trends in the evolution of the world securities market can be distinguished:

1) Introduction of information technologies.

2) The growth of the capitalization level

3) The increase in the level of regularity and organization

4) Securitization

5) Formation of the global stock market

6) Centralization and concentration of capital.

The authors (Fang, Yong et al. ,2008), the investor's typology largely determines his investment strategy.

In the opinion to (Toušek, Zdenek et al. ,2022), the risks present on the stock market are associated with the probability of the occurrence of certain adverse circumstances. The authors (Krivko, Mikhail et al. ,2021) identify the following main risks that accompany operations with securities: price risk, portfolio risk, basic risk and exchange rate risk.

Recently, private investors have had more opportunities to work in the stock markets. At the same time, technical and fundamental analysis skills are required for effective work

According to (Krivko, Mikhail et al. ,2021), most private investors act irrationally in the securities market. Often, a private investor makes an analysis of probabilities and complex situations and makes investment decisions based on heuristic principles.

In the opinion to (I Sazonets, et al. ,2021), the efficiency of private investors largely depends on their ability to control their emotions and the ability to navigate the information space, as well as the ability to use technical and fundamental methods.

In views to (Yekimov S., et al.,2022), the decision regarding the sale or purchase of securities should be made taking into account the dynamics of the liquidity of securities and the study of the risks associated with these securities.

The authors (Rumánková, Lenka et al. 2019), the attractiveness of the stock market for private investors largely depends on its information efficiency.

Weak information efficiency makes it difficult to predict the future price of a security, and therefore complicates the decision to buy and sell a security based on technical analysis. At the same time, the availability of high information efficiency of the stock market contributes to attracting private traders, including for financial investments in the medium and long term.

Evaluation of the choice of the investment portfolio structure based on three parameters: profitability, risk and correlation faces the problem of measurement accuracy these values . If the error size is large enough, then a mathematical model based on such data is not suitable for practical use. In this regard, the development of alternative economic and mathematical models tosolve the problem of forming an optimal investment portfolio deserves attention.

Recently, the Capital Asset Pricing Model proposed by (Sharpe, William F,1964) has been widely applied in practice.

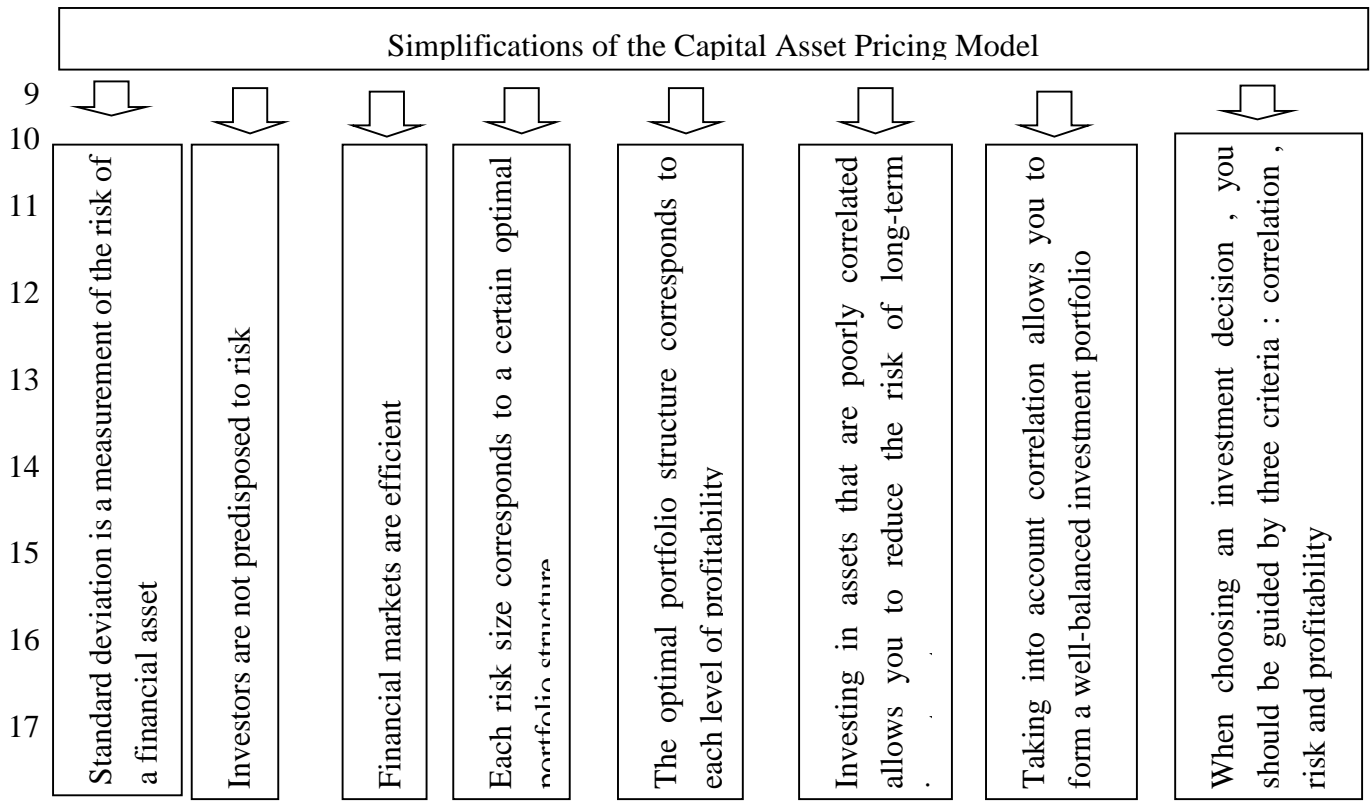

**Figure 2.** Simplifications of the Capital Asset Pricing Model

According to this model, the premium for the existing risk varies in proportion to the coefficient β of the investment portfolio or security.

$$E(R_i) = R_f + \beta_i(E(R_m) - R_f) \qquad (1)$$

where $E(R_i)$ -- is the expected return on investment activity.

$R_f$ — the rate of return on investing in a risk-free asset

$\beta_i$ — the coefficient determining the degree of sensitivity the asset to changes in the value of market profitability.

$E(R_m)$ - expected rate of return of the market portfolio

$E(R_m) - R_f$ - risk premium from investment activity

$$\beta_i = \frac{cov(R_i, R_m)}{\sigma^2(R_m)} \qquad (2)$$

According to a number of authors (Bhutta, Nousheen et al.,2020), the Capital Asset Pricing Model contains a number of simplifications (Figure 2), which on the one hand simplify its practical use, and on the other hand makes it difficult to apply correct practical solutions based on this model.

The authors to (Kolari, James & Liu, Wei & Huang, Jianhua,2021), the expected return is the main parameter in portfolio theory, but its objective definition is difficult.

In the opinion to (Aldabbagh, Harith et al. ,2022), a problem for the practical use of the Capital Assets Pricing Model in some cases may be the presence of a large error in calculating correlation, covariance, average yield and variation.

In views to (Luca, Pasquale, 2022), historical data on the basis of which the main parameters of portfolio theory are determined can be a tool for predicting their future changes.

In this regard, the development of other theoretical approaches to determine the optimal structure of the investment portfolio deserves attention.

## 2. Methodology

Orthogonal polynomials on the orthogonality interval $(a, b) \to R$ called (Shoukralla, Emil,2021) infinite sequence of real polynomials $p_0(x), p_1(x), p_2(x), \ldots, p_n(x), \ldots$ where is each of the polynomials $p_n(x)$ has a degree $n$, which are mutually orthogonal in the sense of a scalar product defined in space $L^2$.

The system of orthogonal polynomials $p_n(x)$, is complete. This means that any polynomial $F(x)$ degrees $n$ on the orthogonality interval $(a, b) \to R$ can be written as a series:

$$F(x) = \sum_{i=1}^{n} c_i p_i(x) \qquad (3)$$

where $c_i$ - decomposition coefficients

A variety of orthogonal polynomials are Chebyshev polynomials of the first kind $T_n(x)$, they can be written in general form (Anshelevich, Michael,2016):

$$T_n(x) = \cos(n \cdot \arccos(x)) \qquad (4)$$

For Chebyshev polynomials of the first kind $T_n(x)$ there is a recurrent formula:

$$T_{n+1}(x) = 2xT_n(x) - T_{n-1}(x) \qquad (5)$$

Chebyshev polynomials of the first kind are used in practice to approximate the empirical data obtained by a function (Guan, W.J. et al. ,2018). Thus , the dynamics of the exchange value of a security in a given time interval $[t_0, t_1]$, The information obtained in the form of a numerical series from statistical sources can be represented as an expansion over the orthogonal Chebyshev polynomials of the first year with some given accuracy.

$$Q(t) = \sum_{i=1}^{n} c_i T_i(t) \qquad (6)$$

Где $Q(t)$ − dependence of the exchange value of a security on time, $c_i$ – decomposition coefficients, $T_n(x)$ - Chebyshev polynomials of the first kind.

Thus, for each j-th paper included in the investment portfolio, it is possible to determine the representation of its exchange value in the form of a decomposition by orthogonal Chebyshev polynomials of the first kind

$$Q^j(t) = \sum_{i=1}^{n} c_i^j T_i^j(t) \qquad (7)$$

Under the scalar product of vectors on $n-$ dimensional euclidean space $\boldsymbol{a} = (a_1, a_2, \ldots, a_n)$ and $\boldsymbol{b} = (b_1, b_2, \ldots, b_n)$ understood (Kim, 2018)

$$\langle \boldsymbol{a}, \boldsymbol{b} \rangle = a_1 b_1 + a_2 b_2 + \cdots + a_n b_n \qquad (8)$$

Expression (8) is equivalent to

$$\langle \boldsymbol{a}, \boldsymbol{b} \rangle = |\boldsymbol{a}||\boldsymbol{b}|\cos(\theta) \qquad (9)$$

We generalize expression (9) for vectors of the space formed by orthogonal Chebyshev polynomials of the first kind

$$\langle Q^i(t) Q^j(t) \rangle = |Q^i(t)||Q^j(t)|\cos(\theta) \qquad (10)$$

$$\cos(\theta)_{ij} = \frac{\sum_{k=1}^{n} c_k^i c_k^j}{\sqrt{[\sum_{k=1}^{n}(c_i^2)]}\sqrt{[\sum_{k=1}^{n}(c_j^2)]}} \qquad (11)$$

where $\cos(\theta)_{ij}$ this is the cosine of the angle between the vectors in the space formed by the orthogonal Chebyshev polynomials of the first kind.

## 3. Results

Within the framework of this study, the author analyzed the dynamics of the exchange value of eight Czech joint-stock companies: CEZ as (CEZP), Colt CZ Group SE (CZG), Erste Bank (ERST), Komercni Banka (BKOM), Moneta Money Bank AS (MONET), Photon (PENP), Vienna insurance group (VIGR)

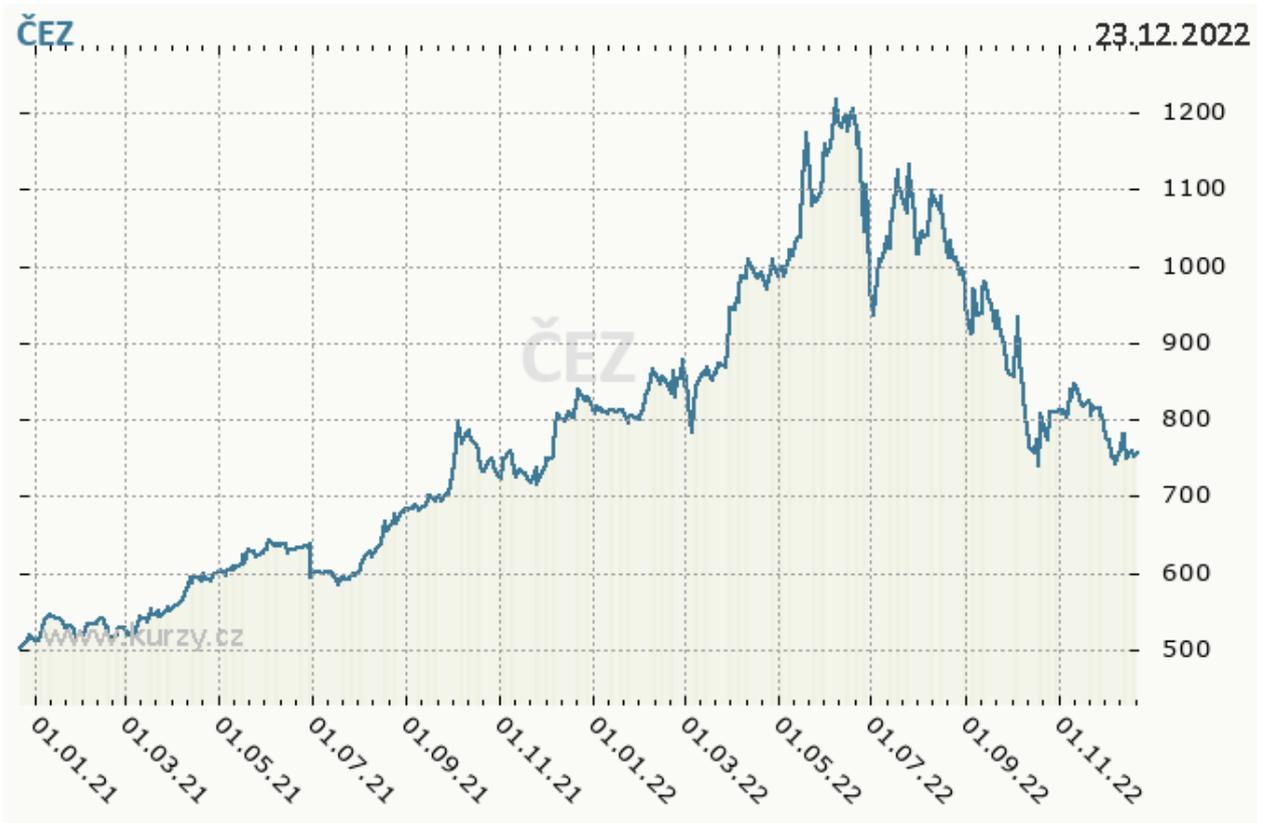

59

**Figure 3.** Dynamics of CEZ as (CELP) shares in 2022 ( website data www.kurzy.cz )

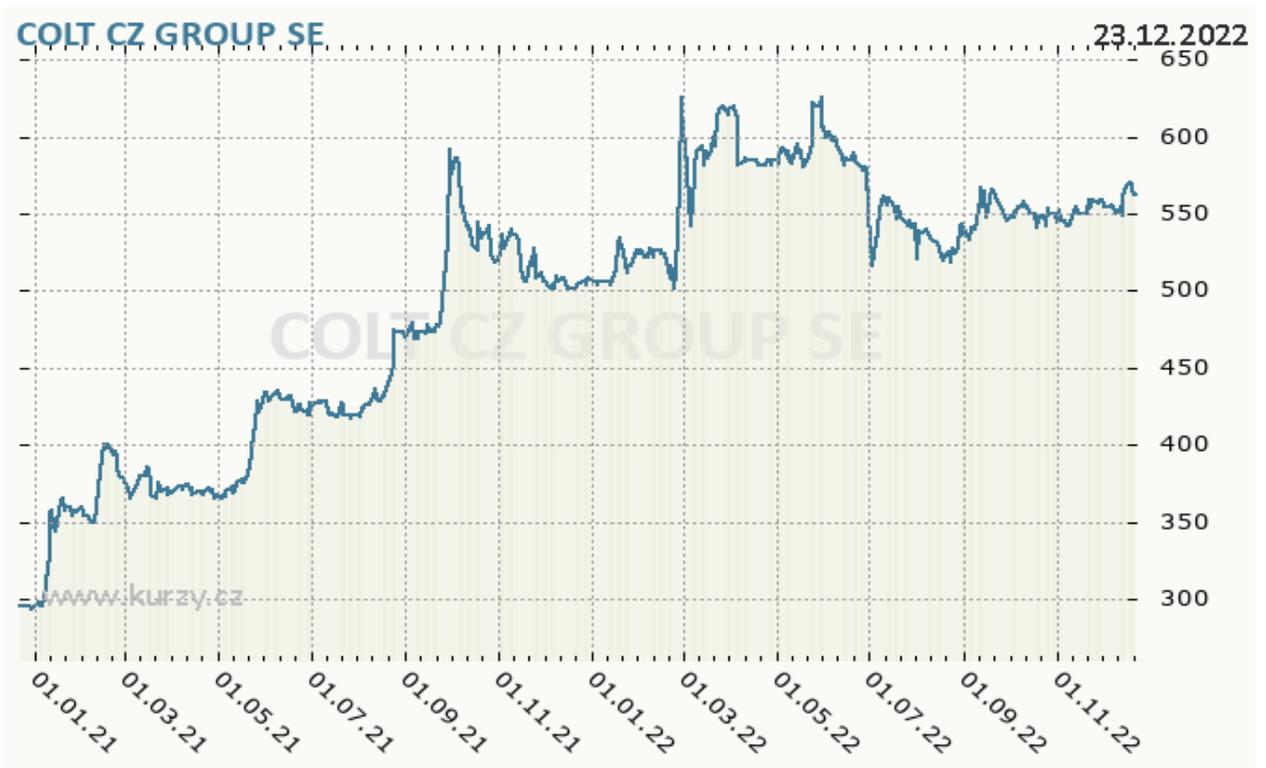

60

**Figure 4.** Dynamics of shares of Colt CZ Group SE (CZ) in 2022(website data www.kurzy.cz)

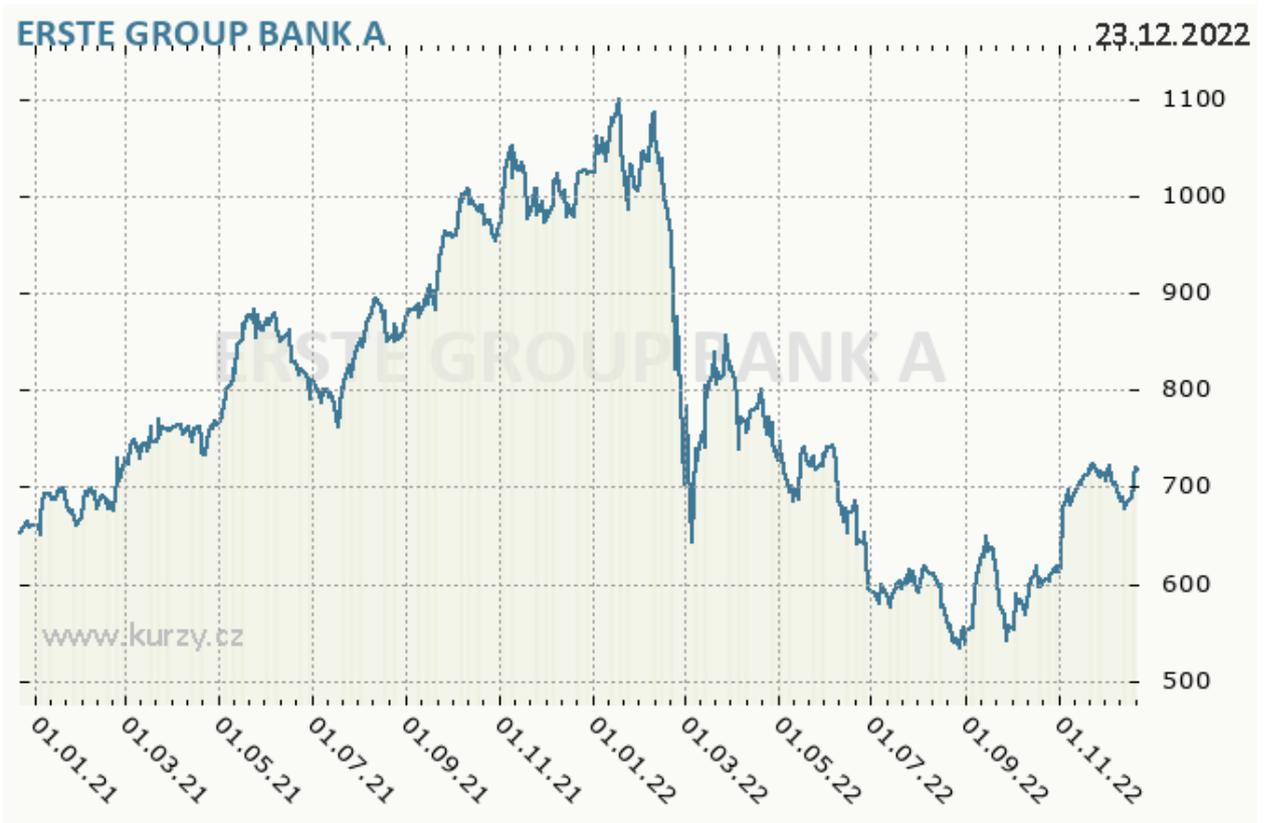

**Figure 5.** Dynamics of Erste Bank (ERSTE) shares in 2022 (website data www.kurzy.cz )

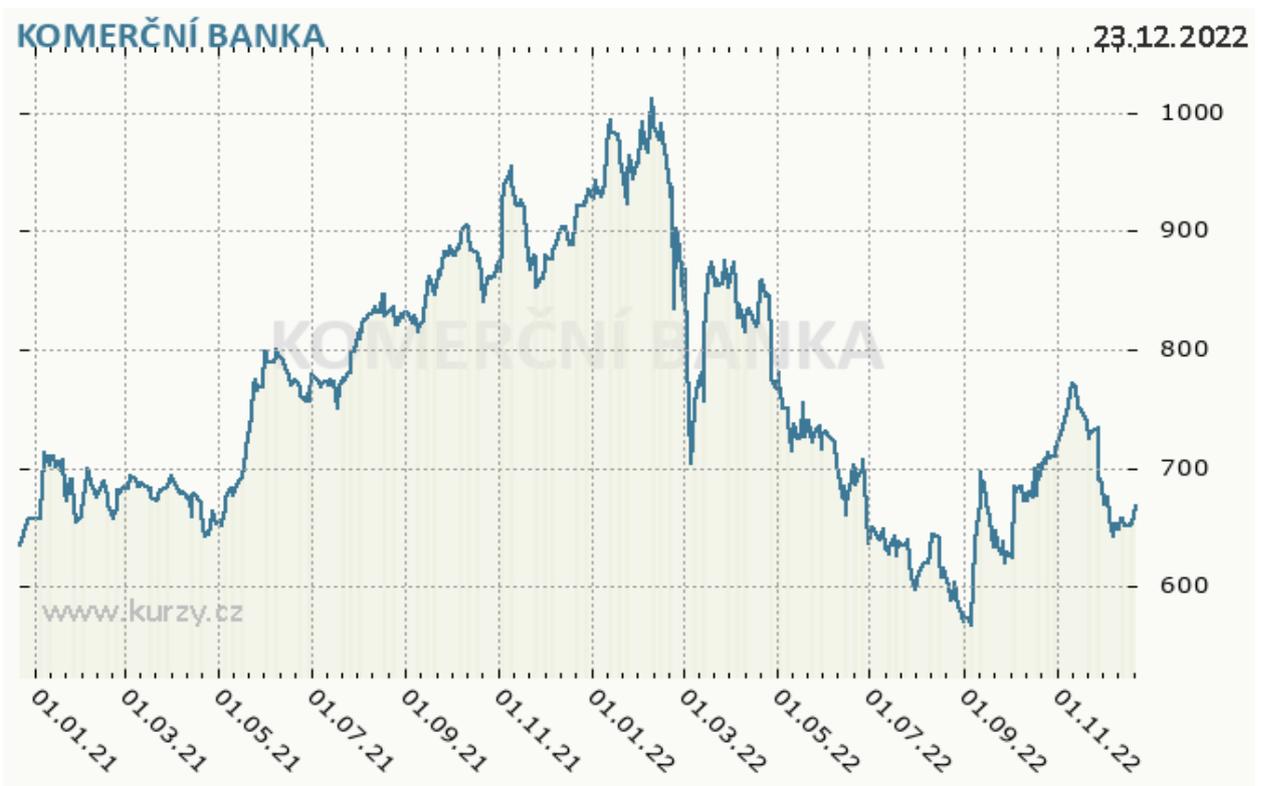

**Figure 6.** Dynamics of Komercni Banka (COM) shares in 2022 (website data www.kurzy.cz )

64

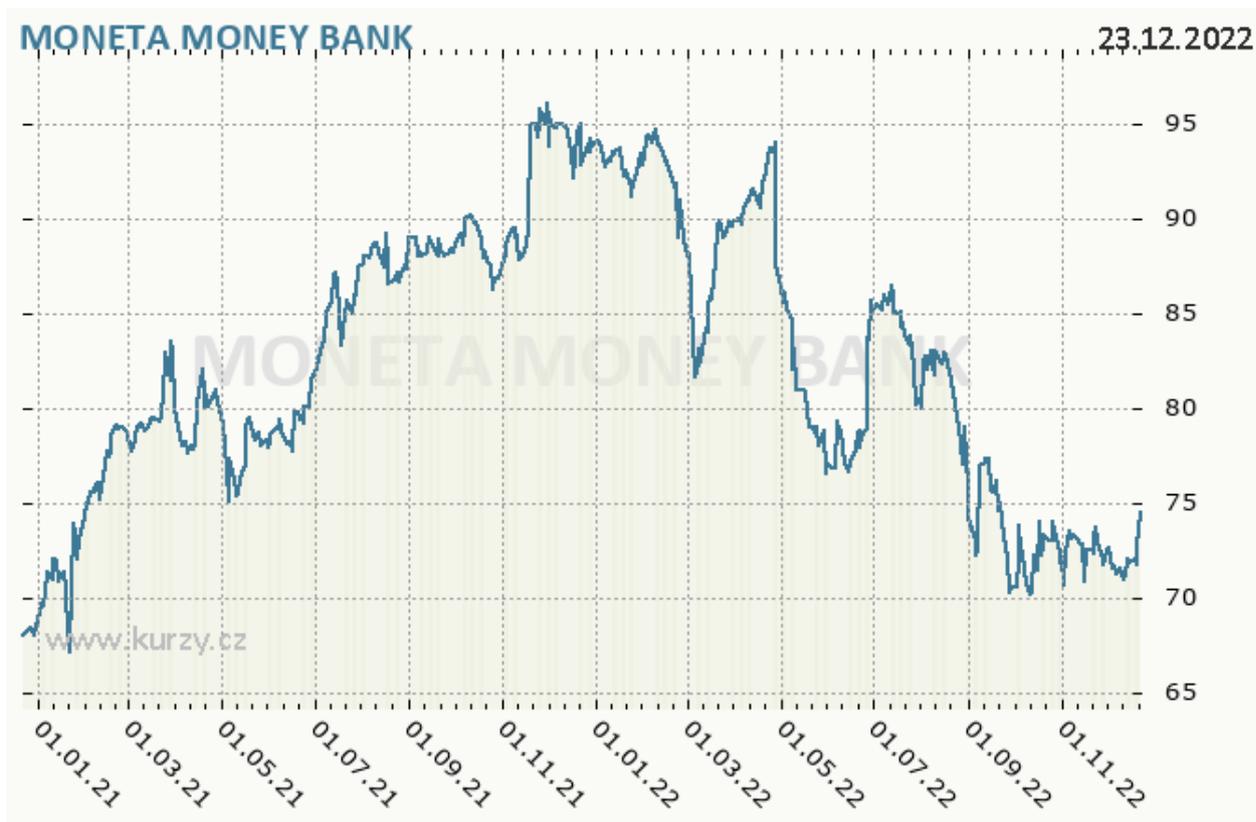

**Figure 7.** Dynamics of Moneta Money Bank AS (MONEY) shares in 2022 (website data www.kurzy.cz )

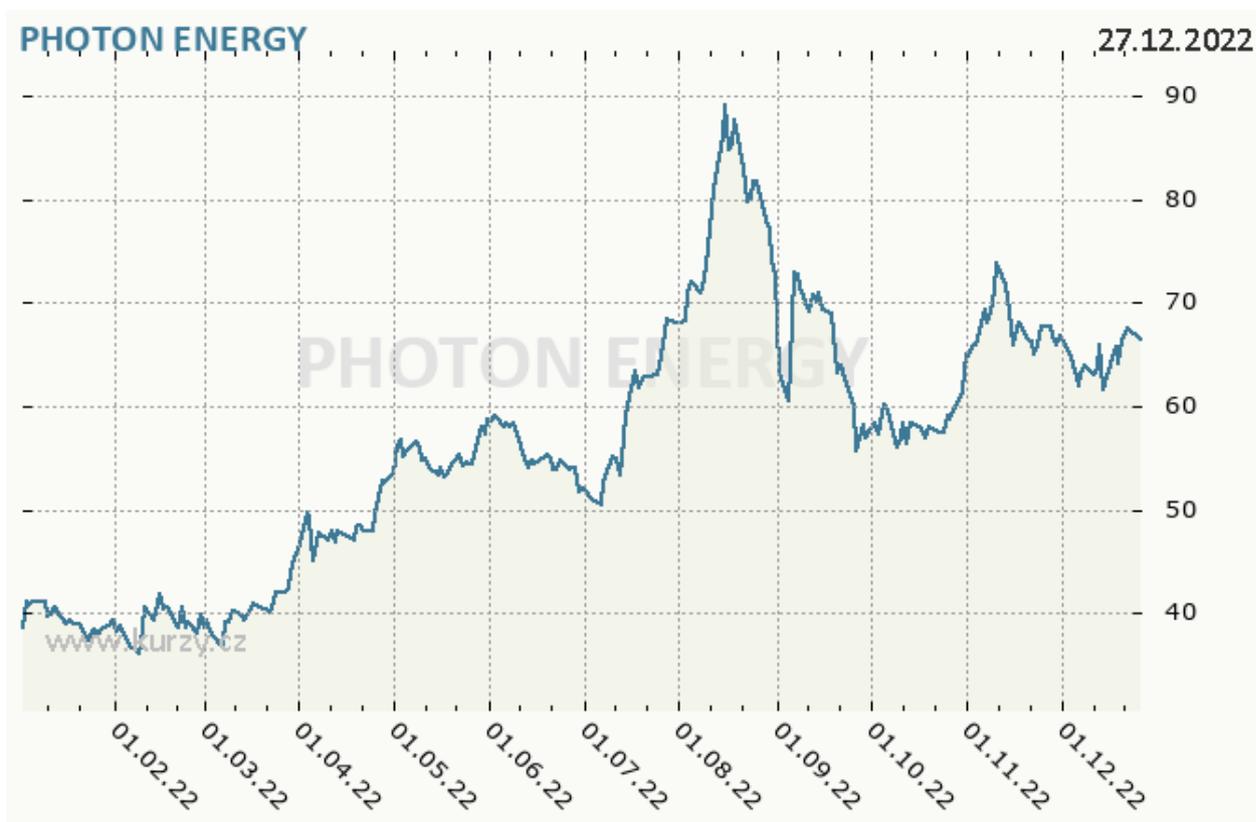

65

**Figure 8.** Dynamics of Proton (PUMP) shares in 2022 (website data www.kurzy.cz)

66

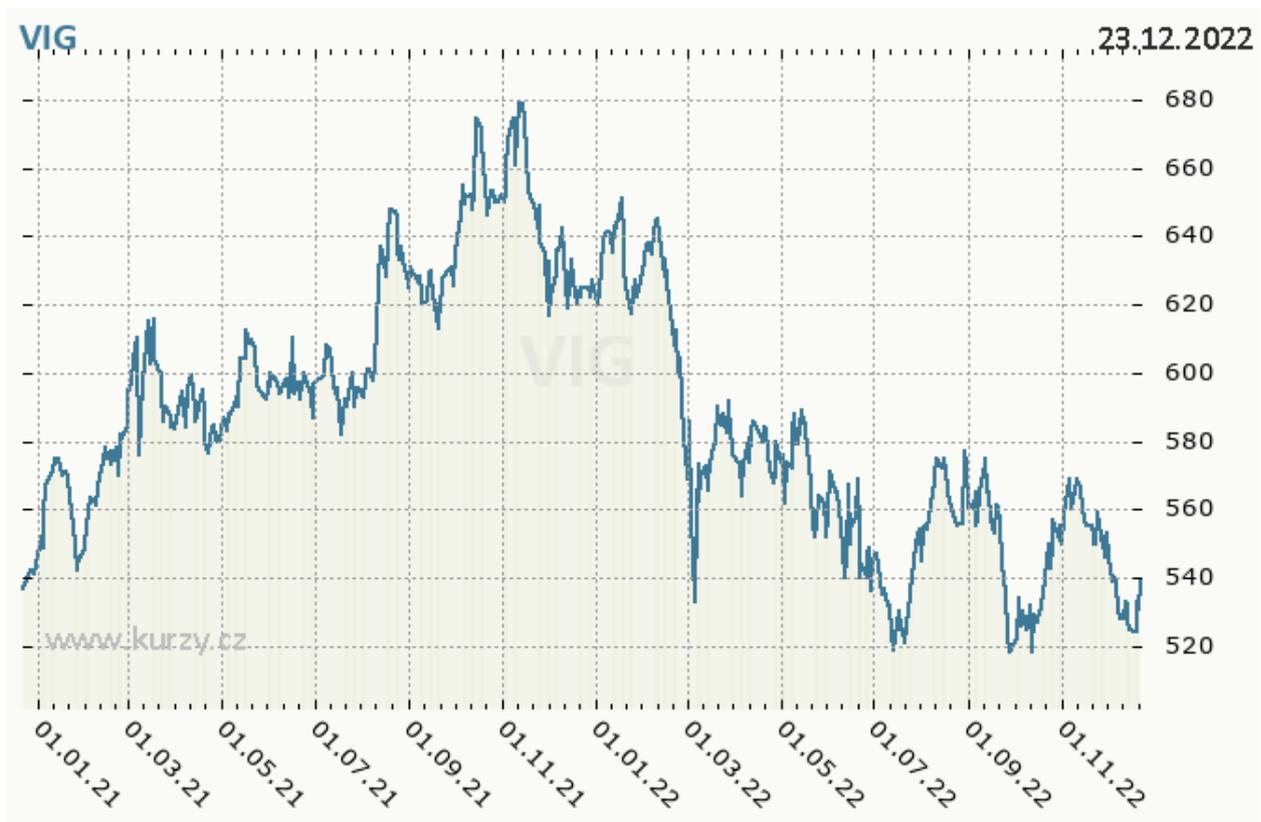

**Figure 9.** Dynamics of Vienna insurance group (VIG) shares in 2022 ( website data www.kurzy.cz )

In our opinion, an investor , when making a decision to purchase a security, is guided, in many respects, by an heuristic approach . The concepts of variance and correlation are not observable by human senses , but at the same time, even a person far from investment activity will always be able to say that the images on the two charts are different.

As part of this study, using the MATHLAB software, the author decomposed the time series of the exchange rate value of shares of eight Czech companies in 2021. The number of terms of the expansion (7) by Chebyshev polynomials of the first kind was 227.

The resulting vectors were normalized by one, and then the cosines of the angles between these vectors (11) were calculated.

The calculation results are shown in Table 1

**Table 1.** Table of cosines of angles (11) between the time series expansion vectors of the exchange rate value of shares of Komercni Banka (BKOM), CEZ A.S. (CEZP), Colt CZ Group SE (CZG), Erste Bank (ERST), Kofola CeskoSlovensko A.S., Moneta Money Bank A.S. (MONET), Photon (PENP), Vienna insurance group (VIGR) on Chebyshev polynomials of the first kind, obtained after normalization of vectors (6) by one, (obtained by the author).

| $\cos(\theta)_{ij}$ | Komercni Banka (BKOM) | CEZ a.s. (CEZP) | Colt CZ Group SE (CZG) | Erste Bank (ERST) | Kofola CeskoSlovensko AS | Moneta Money Bank AS (MONET) | Photon (PENP) | Vienna insurance group (VIGR) |
|---|---|---|---|---|---|---|---|---|
| Komercni Banka (BKOM) | 1.000 | 0.949 | 0.977 | 0.996 | 0.983 | 0.989 | 0.890 | 0.988 |
| CEZ AS (CEZP) | 0.949 | 1.000 | 0.984 | 0.937 | 0.985 | 0.977 | 0.951 | 0.978 |
| Colt CZ Group SE (CZG) | 0.977 | 0.984 | 1.000 | 0.965 | 0.990 | 0.990 | 0.950 | 0.994 |
| Erste Bank (ERST) | 0.996 | 0.937 | 0.965 | 1.000 | 0.976 | 0.981 | 0.874 | 0.981 |
| Kofola CeskoSlovensko AS | 0.983 | 0.985 | 0.990 | 0.976 | 1.000 | 0.997 | 0.927 | 0.994 |
| Photon (PENP) | 0.890 | 0.951 | 0.950 | 0.874 | 0.927 | 0.926 | 1.000 | 0.944 |
| Vienna insurance group (VIGR) | 0.988 | 0.978 | 0.994 | 0.981 | 0.994 | 0.996 | 0.944 | 1.000 |

In parallel with this, correlation coefficients were calculated between time series of exchange rates of shares of Komercni Banka (BKOM), CEZ A.S. (CEZP), Colt CZ Group SE (CZG), Erste Bank (ERST), Kofola CeskoSlovensko A.S., Moneta Money Bank A.S. (MONET), Photon (PENP), Vienna insurance group (VIGR)

**Table 2.** Table of correlation coefficients cov(R_i,R_j ) between time series of exchange rates of shares Commercial Bank (BKOM), CEZ as (CEZP), Colt CZ Group SE (CZG), Erste Bank (ERST), Kofola CeskoSlovensko as, Moneta Money Bank AS (MONET), Photon (PENP), Vienna insurance group (VIGR), (obtained by the author).

| $cov(R_i, R_j)$ | Komercni Banka (BKOM) | CEZ a.s. (CEZP) | Colt CZ Group SE (CZG) | Erste Bank (ERST) | Kofola CeskoSlovensko AS | Moneta Money Bank AS (MONET) | Photon (PENP) | Vienna insurance group (VIGR) |
|---|---|---|---|---|---|---|---|---|
| Komercni Banka (BKOM) | 1.000 | -0.296 | 0.069 | 0.940 | 0.482 | 0.699 | -0.782 | 0.821 |
| CEZ AS (CEZP) | -0.296 | 1.000 | 0.332 | -0.270 | 0.499 | 0.165 | 0.200 | -0.079 |
| Colt CZ Group SE (CZG) | 0.069 | 0.332 | 1.000 | -0.010 | 0.248 | 0.038 | -0.265 | -0.140 |
| Erste Bank (ERST) | 0.940 | -0.270 | -0.010 | 1.000 | 0.499 | 0.661 | -0.713 | 0.842 |
| Kofola CeskoSlovensko AS | 0.482 | 0.499 | 0.248 | 0.499 | 1.000 | 0.852 | -0.541 | 0.611 |
| Photon (PENP) | 0.699 | 0.165 | 0.038 | 0.661 | 0.852 | 1.000 | -0.630 | 0.712 |
| Vienna insurance group (VIGR) | -0.782 | 0.200 | -0.265 | -0.713 | -0.541 | -0.630 | 1.000 | -0.507 |

Further , the correlation coefficients between those presented in Tables 5 and 6 were determined $\cos(\theta)_{ij}$ and $cov(R_i, R_j)$

$cov(cov(R_i, R_j), \cos(\theta)_{ij})$ =0.830724856

From this, the author concludes that the use of $\cos(\theta)_{ij}$ along with $cov(R_i, R_j)$ it may be of practical importance, due to their high mutual correlation.

The correlation indicator $\cos(\theta)_{ij}$ and $cov(R_i, R_j)$ is equal 0.830724856 , which means that both models can give similar forecasts regarding the formation of the optimal structure of the investment portfolio, but the model proposed in this paper is a different mathematical apparatus based on Chebyshev polynomials of the first kind. Which are quite widely used to solve technical problems when analyzing the spectrum of audio signals.

## 4. Discussion

Chebyshev polynomials of the first kind have long been used to approximate experimental data when solving various technical problems. Within the framework of this study, the dynamics of 8 shares of Czech enterprises was analyzed by means of decomposition by Chebyshev polynomials. The vectors of decomposition of time series of exchange values of securities allow to analyze the dynamics of exchange values of securities more effectively if their dynamics does not correspond to the normal distribution law.

By analogy with (1) in this paper, the author suggests using to determine the expected $E(R_i)$ return on investment activity model

$$E(R_i) = R_f + \varphi(E(R_m) - R_f \qquad (12)$$

where $E(R_i)$- is the expected return on investment activity.

$R_f$ — the rate of return on investing in a risk-free asset

$\beta_i$ — the coefficient determining the degree of sensitivity $i$ — the asset to changes in the value of market profitability $R_m$.

$E(R_m)$ - expected rate of return of the market portfolio

$E(R_m) - R_f$ - risk premium from investment activity

$\varphi = \cos(\theta)_{im}$ - is the cosine of the angle (11) between the vectors in the space formed by

the orthogonal Chebyshev polynomials of the first year for the market and for a certain security. For the securities market $\varphi = 1$, each type $i$ of securities will have its own individual coefficient $\varphi_i$. The economic meaning of the indicator $\varphi_i$ the question is how volatile this type of security is compared to the market.

By meaning $-1 \leq \cos(\theta)_{im} \leq 1$, in this, it resembles the indicator from the Capital Asset Pricing Model. However, the statistical functions variance and correlation are not used in the calculation.

In our opinion, an investor, when making a decision to purchase a security, is guided largely by a heuristic approach. And correlation and variance are not observed by human senses.

## 5. Conclusions

The theory proposed by the author allows us to determine how interconnected the dynamics of the exchange value of securities and , but for this the author uses instead of the correlation coefficient

$$cov(R_i, R_j) = \frac{1}{n}\sum_{t=1}^{n}(R_i^t - \bar{R}_i)(R_j^t - \bar{R}_j) \qquad (13)$$

The cosine of the angle between the vectors formed as a result of the decomposition of the statistical series of securities yields by orthagonal Chebyshev polynomials of the first kind (11).

$$P_N(x) = \sum_{k=0}^{N} c_k T_k(x) = c_0 T_0(x) + c_1 T_1(x) + \cdots + c_N T_N(x) \qquad (14)$$

Coefficient from the model (12) $\varphi = \cos(\theta)_{im}$ it shows how variable the profitability of the i-th financial asset is compared to the profitability of the market.

Within the framework of this study, statistical data of the largest Czech companies whose shares are listed on the Prague Stock Exchange were analyzed. The industry affiliation of these companies is not of fundamental importance for the model proposed in this paper.